\documentclass[12pt]{article}
\usepackage[centertags]{amsmath}
\newcommand{\la}{\lambda}
\begin{document}

\title{The time factor in the semi-classical approach to the Hawking radiation}
\date{submitted 30/12/2008}
\author{M. Pizzi\\Physics Department, University of Rome ``La Sapienza'', \\P.le A. Moro, Italy 00185, \\E-mail: pizzi@icra.it}

\maketitle
\begin{abstract}
The Hawking radiation in the semi-classical approach is re-considered. In the so-called ``Angheben method'' for the calculation of the imaginary part of the action it was missed the temporal part contribution. This has been recently noticed by Akhmedov et \textsl{al.}, but also there the time part was not properly considered, the sign being reversed. We show that using the semi-classical approach on a fixed background it is not possible to find any tunneling effect from the interior to the exterior of the Schwarzschild black hole. The same critic applies to the derivations which use the Painlev$\acute{e}$ coordinates: also in this procedure it was missed the temporal factor contribution. In this way it is naturally solved also the ``factor-two-problem''.
\end{abstract}

\section{Introduction}
\label{sec:Introduction}
How to retrieve the Hawking radiation \cite{Haw} in the semi-classical tunneling picture has been at the center of a recent debate, see e.g. Refs.\cite{AAS,APS,APdS,Ang,NVZ,KM,KW,PW00}. Our claim is that this tunneling is not allowed.

Recently, Akhmedov et \textsl{al.} \cite{APS} have noted that in Ref.\cite{Ang} the temporal part contribution in the tunneling was overlooked. However this was already noted by Belinski (see \cite{Bel07}, footnote 1): the derivation given in Ref.\cite{Ang}, as well as all the others which follows the same path, e.g. Refs.\cite{AAS,NVZ,KM}, considers only the imaginary contribution of the function $W(r)=\int p_r dr$, and not of the whole action which is $S=-Et+W(r)$. Such procedure is good in non-relativistic quantum mechanics, but in the black hole space-time also the variable $t$ acquires an imaginary part crossing the horizon.
This factor $-Et(r)$, if properly considered, cancels exactly the imaginary part acquired by $W(r)$ \cite{Bel07}. Also the earlier work of Parikh and Wilzcek \cite{PW00}, which uses the Painlev$\acute{e}$ coordinates, is affected by the same overlooked factor. The correction proposed in Ref.\cite{APS} has the wrong sign. 
The ground of the semi-classical method is re-called in Appendix 1; the massless case is briefly considered in Appendix 2.

%________________________________________________________________________________________________________
\subsection{Tunneling in Schwarzschild coordinates; the simplest way}
 Let us take the Schwarzschild metric (it is sufficient to consider just the radial part):
\begin{eqnarray}\label{1.1}
	 ds^{2}=f(r) dt^{2}-f^{-1}(r)dr^{2}\,, \\
	 f(r)=1-\frac{r_H}{r}\, ,  \ \ \ \ \ r_H=2M \ .
\end{eqnarray}
 The easiest way to show that the geodesic action does not acquire any imaginary part, is the following: using the two first integrals of the motion
\begin{eqnarray}\label{1.3}
\frac{dt}{ds}=\frac{E}{m\,f(r)}\ ,\ \ \ \ \texttt{and} \ \ \ \  1=f(r)\left(\frac{dt}{ds}\right)^{2}-\frac{1}{f(r)}\left(\frac{dr}{ds}\right)^{2}\ ,
\end{eqnarray} 
one has 
\begin{eqnarray}	 
 \frac{dr}{ds}=\pm \sqrt{\frac{E^{2}}{m^{2}}-f(r)} \ , \label{1.4}
\end{eqnarray} 
(``+'' corresponds to out-going particles, ``--'' to in-going). Then it is immediate to find that the action along the particle's world line is
\begin{equation}\label{1.2}
	S=-m\int ds=-m\int \frac{1}{dr/ds} dr=-m\int \frac{dr}{\pm\sqrt{E^{2}/m^{2}-f(r)}}\ .
\end{equation}
 Thus it is evident that for a particle which can reach infinity\footnote{ Indeed, only these particles could give place to the Hawking radiation.} (i.e. with $E\geq m$), the action (\ref{1.2}) is always real, for any value of the radius $r>0$, and thus also in the vicinity of the horizon (see e.g. \cite{FeyGrav}), therefore:
\begin{equation}\label{1.2b}
	2\texttt{Im}(\Delta S)=\oint_{r_H} \frac{dr}{\mp\sqrt{\frac{E^2}{m^2}-1+\frac{r_H}{r}}}=0.
\end{equation}
The point is that for any trajectory \emph{there is no pole} on the horizon in the action's integrand (contrary to what has been assumed in papers \cite{AAS}-\cite{PW00}). This statement is evident without any further calculation, it is simply due to the invariance of the action and to the fact that any point of the horizon is an ordinary regular space-time point, and so its small vicinity is just an empty flat space-time which can be covered by Minkowski coordinates; therefore no sub-barrier transition creating an imaginary part in the action can appear in this vicinity\footnote{Also using a very different approach (functional Schrodinger formalism) 
some authors arrived to a conclusion compatible with the fact that it can not be any tunnelling through the horizon simply because there is no barrier to penetrate, see \cite{VSK}, Sec.IV.}.

We wish to say that the state of affairs is clear enough that the problem could be considered solved in this way. But since there are several contrasting papers on this subject, in the following we will discuss some of the most frequently used manipulations.
%________________________________________________________________________________________________________

\section{H-J method; Angheben et al. method}
\label{sec:HJMethod}
 Since the Hamiltonian is t-independent, we can use the usual Ansatz: 
\begin{equation}\label{1.5}
	S=-Et+W(r) \ .
\end{equation}
 Then putting it in the Hamilton-Jacobi (H-J) equation
\begin{equation}\label{1.6a}
 \partial_{\mu}S \,\partial^{\mu}S=\frac{1}{f(r)}\left(\frac{\partial S}{\partial t}\right)^2-f(r)\left(\frac{\partial S}{\partial r}\right)^2 =m^2,
\end{equation}
we find
\begin{eqnarray}
\frac{1}{f(r)}E^2-f(r)\left(\frac{dW(r)}{dr}\right)^2 =m^2\nonumber\\
 \Rightarrow  W(r)=\int\frac{\pm\sqrt{E^{2}-m^{2}f(r)}}{f(r)}dr \ . \label{1.6} 
 \end{eqnarray}
Then one has to remember to take into account also the variation of the function $t(r)$ along the trajectory.
The dependence of the time $t$ on the coordinate $r$ (this is the crucial point for the correct calculation of the time contribution)  can be found remembering that along the trajectory $\partial S/\partial E=const.$ (see \cite{LL}), therefore
\begin{eqnarray}
const=-t+\frac{\partial W}{\partial E} \ \ \ \Rightarrow \ \ \ t=\int\frac{\pm E}{f(r)\sqrt{E^{2}-m^2f(r)}}dr \ \label{1.7} 
\end{eqnarray}
(for simplicity we omitted the arbitrary constant, because this term, being constant along the whole trajectory, does not contribute to the $\Delta S$). 
Then substituting the above expressions of $W(r)$ and $t(r)$ in the action (\ref{1.5}), 
we finally have the integral expression of the action:
\begin{equation}\label{1.8}
 S=-\int \frac{\pm E^{2}}{f(r)\sqrt{E^{2}-m^{2}f(r)}}dr+\int\frac{\pm\sqrt{E^{2}-m^{2}f(r)}}{f(r)}dr,
\end{equation}
which gives exactly Eqn.(\ref{1.2}). Therefore, even if one wants to preserve the splitting of Eqn.(\ref{1.8}), it is easy to see that the imaginary part acquired by the second integral term  on the horizon $r_H$,
\begin{equation}\label{1.9}
	\Delta W=\frac{1}{2}\oint_{r_H}\frac{\pm\sqrt{E^{2}-m^{2}f(r)}}{f(r)}dr=\pm i2\pi M E \ \ \ \ \textit{(Angheben et al.)},
\end{equation}
is exactly canceled by the temporal part of the action (which has been overlooked in \cite{Ang} and in many other papers) which is:
\begin{equation}\label{forgotten}
	-E\Delta t=-\frac{1}{2}\oint_{r_H}\frac{E^{2}}{\pm f(r)\sqrt{E^{2}-m^{2}f(r)}}dr=\mp i2\pi M E \ \ \ \ \textit{(overlooked term)}.
\end{equation}
The sum of these two terms is clearly zero, in accordance with our previous result, Eqn.(\ref{1.2b}).

\section{Akhmedov et al. correction}
\label{sec:AkhmedovEtAlCorrection}

In Ref.\cite{APS} (and in other following papers) the temporal factor $-E\Delta t$ is not forgotten, but it is yet considered in an incorrect way, the problem being in the sign. Their argument is that, considering the Kruskal-Szekers (KS) coordinates, 
\begin{eqnarray}
\left\{\begin{array}{l}	
T= e^{r/2r_H}\sqrt{\frac{r}{r_H}-1}\   \sinh \left(\frac{t}{2r_H}\right) \\
R= e^{r/2r_H}\sqrt{\frac{r}{r_H}-1}\   \cosh \left(\frac{t}{2r_H}\right)
\end{array}\right. & r>r_H \label{k1} \\
\left\{\begin{array}{l}
T= e^{r/2r_H}\sqrt{1-\frac{r}{r_H}}\  \cosh \left(\frac{t}{2r_H}\right)\\ 
R= e^{r/2r_H}\sqrt{1-\frac{r}{r_H}}\ \sinh\left(\frac{t}{2r_H}\right)
\end{array}\right. & r<r_H  \label{k2} \ ,
\end{eqnarray}
then, they say, in order to cross the horizon the Schwarzschild time ``t'' should  be changed with $t \rightarrow t-i \pi 2M$.
Indeed this term, ignoring the sign, corresponds to our integral (\ref{forgotten}). 

However, using the same reasoning it is well possible to make the opposite ``rotation'', i.e. $t \rightarrow t+i \pi 2M$. In this case the $+i$ can be absorbed by the square root in this way
\begin{eqnarray}
T_{out}= e^{r/2r_H}\sqrt{\frac{r}{r_H}-1}\ \sinh \left(\frac{t}{2r_H}\right) \nonumber&\\
\stackrel{t\rightarrow t+i\pi r_H}{\longrightarrow} &  e^{r/2r_H}\sqrt{\frac{r}{r_H}-1}\  \sinh \left(\frac{t}{2r_H}+i\frac{\pi}{2}\right)\nonumber\\
=& e^{r/2r_H}\sqrt{\frac{r}{r_H}-1}\  (+i)\cosh \left(\frac{t}{2r_H}\right)\nonumber\\
=& e^{r/2r_H}\sqrt{\left(\frac{r}{r_H}-1\right)(-1)}\  \cosh \left(\frac{t}{2r_H}\right) \label{3.12} \nonumber\\
=&T_{in}\ ,
\label{3.13} 
\end{eqnarray}
(note that this ``trick'' is possible because the square root is a polydromic function); the same argument holds for the $R$ coordinate. Now the point is that the prescription used by Akhmedov \textsl{et al.} evidently corresponds, in our framework, to take a clockwise path in the calculation of $-E\Delta t$ [see Eqn.(\ref{forgotten})]. However the Cauchy formula used in the integral $\Delta W$ (as well as in $S_0$, in \cite{APS}'s notation) implicates that the path is \emph{counterclockwise}, and for coherence one has to chose the same prescription in both the integrals; when the signs are coherently calculated the sum of $-E\Delta t+\Delta W$ is zero.

It is easy to see that the action has no discontinuities on the horizon along the geodesics also in KS coordinates, in the following way. The metric now is
\begin{eqnarray}\label{ds2}
 ds^2=\omega(r)(dT^2-dR^2)\ , \ \ \ \  \omega(r)\equiv\frac{4r_H^3}{r}e^{-r/r_H} \ ,
\end{eqnarray}
where the implicit connection with the radial Schwarzschild coordinate is given by
\begin{eqnarray}
 \left(\frac{r}{r_H}-1\right)e^{r/r_H}=R^2-T^2 \ .\label{KS1}
\end{eqnarray}
The action along the geodesic is proportional to the proper time of the particle
\begin{equation}\label{B1}
	S=-m\int ds \ .
\end{equation}
The geodesic equations have a first integral that can be written in KS coordinates as follows \cite{Bel06}:
\begin{eqnarray}
 \omega(r)\left(T\frac{dR}{ds}-R\frac{dT}{ds}\right)&=&\frac{2r_H E}{m} \ ,\label{B3}
\end{eqnarray}
where $E$ is the Schwarzschild conserved energy of the particle. From (\ref{ds2}) and (\ref{B3}) one can find the geodesic trajectory $T(R)$. Due to the analyticity of $\omega(r)$ on the horizon points, the function $T(R)$ will be also analytical at the horizon points (that is at the points where $T=R$). Then from (\ref{B3}) follows that the differential $ds$ written in the form 
\begin{eqnarray}\label{B4}
 ds=\frac{m}{2r_HE}\omega(r)\left[T(R)-R\frac{dT(R)}{dR}\right]dR  
\end{eqnarray}
gives the integrand for (\ref{B1}) which is also analytical (i.e. with the coefficient in front of $dR$ having no singularities) at the horizon points. This means that nothing special happens to the action when the particle trajectory crosses the horizon and $S$ is a real function without discontinuities on the horizon, therefore one finds $\texttt{Im}[\Delta S]=0$ also in these coordinates. Obviously the result obtained from the above discussion is not surprising since the action is an invariant.

An explicit and detailed treatment of quantum tunneling in KS coordinates can be found in Ref.\cite{Bel06}, where the conclusion is indeed that the tunneling through the horizon is not possible. 
%---------------------------------------------------------------------------------------------------------------------
\section{Painlev$\acute{e}$ coordinates}
\label{Painlev}
 We repeat the same procedure in the frequently-used Painlev$\acute{e}$ coordinates (see e.g.\cite{KW,PW00}).
  By the transformation
\begin{equation}\label{P.1}
\left\{\begin{array}{l}
	 t_P=t+2\sqrt{2M r}+2M \ln  \frac{\sqrt{r}-\sqrt{2M}}{\sqrt{r}+\sqrt{2M}}
	 \\
	 r_P=r
\end{array}\right.,
\end{equation}
 we find the new expression of the line element,
\begin{equation}\label{P.2}
\begin{array}{l}
	 ds^{2}=(1-\la^2) dt_P^{2}-2 \la dt_Pdr- dr^2 \ ,\ \ \la=\sqrt{\frac{2M}{r}}.
\end{array}
\end{equation}
Thus, using the usual Ansatz $S=-Et_P+W_P(r)$, the H-J equation gives:
\begin{equation}\label{P.3}
E^2+2E\la \frac{\partial W_P}{\partial r}-(1-\la^2)\left(\frac{\partial W_P}{\partial r}\right)^2 =m^2,
\end{equation}
therefore
\begin{equation}\label{P.4}
W_P(r)=\int \frac{-E\la\pm\sqrt{E^2+m^2(\la^2-1)}}{\la^2-1} dr.
\end{equation}
Again, the Painleve-time $t_P$ is linked to $r$ by $\partial S/\partial E=const.$, and thus:
\begin{equation}\label{P.5}
t_P(r)=\int\left[-\la\pm \frac{E}{\sqrt{E^2+m^2(\la^2-1)}}\right] \frac{dr}{\la^2-1}.
\end{equation}
Therefore the complete action is:
\begin{eqnarray}
S&=-E\int\left[-\la\pm \frac{E}{\sqrt{E^2+m^2(\la^2-1)}}\right] \frac{dr}{\la^2-1}+\int \frac{-E\la\pm\sqrt{E^2+m^2(\la^2-1)}}{\la^2-1} dr \nonumber \\
&= \pm\int \frac{m^2}{\sqrt{E^2+m^2(\la^2-1)}}dr \ , \label{P.6}
\end{eqnarray}
which is equal to Eqn.(\ref{1.2}), as we expected.
%--------------------------------------------------------------------------------------------------------------------

\section{Conclusions}
\label{sec:Conclusions}
 
 We have shown that it is impossible to achieve the Hawking radiation in the semi-classical tunneling picture. The derivations present in literature are affected by a not-allowed ``splitting'' of the action: they consider only the radial part $W(r)$ of the action, neglecting (or considering with the reversed sign) the imaginary contribution which comes from the time coordinate. Also the discrepancy in the ``factor two'' which is found by some authors using different coordinates systems is simply due to this not-proper treatment of the time: $W(r)$ alone indeed is not invariant; while the whole action, if correctly considered, gives \emph{invariantly} zero for the imaginary part.
  
 Although the tunneling is impossible on a fixed background, what happens if the back-reaction is taken in consideration in a proper way, we do not know yet (in any case it is clear that to try to get a result from the pole singularity on horizon, no matter the back-reaction, is evidently an ill-conceived enterprise). However this is a different problem, and also in the Hawking original derivation the background was assumed to be fixed. Therefore the impossibility of the semi-classical tunneling brings heavily into question also the correctness of the original Hawking derivation; this is in complete agreement with the conclusions of Ref.\cite{Bel06}. 
 
%%%%%%%%%%%%%%%%%%%%%%%%%%%%%%%%%%%%%%%%%%%%%%%%%%%%%%%%%%%%%%%%%%%%%%%%%%%%%%%%%%%%%%%%%%%%%%%%%%%%%%%%%%%%%%%%%%%%%%
\section*{Appendix 1: Semi-classical tunneling}
Just for completeness we recall that the probability of a pair creation during the history of a field is 
\cite{EH,Sch,Pop}  
\begin{eqnarray}
\Gamma=1-e^{-2|\texttt{Im}(\Delta S)|} \ ,
\end{eqnarray}
where $\texttt{Im}(\Delta S)$ is the imaginary part of the field's action acquired in the complex path between the space-time points in which the two particles are created; the main contribution is given just by the first loop (see \cite{Pop}). The semi-classical approach consists in considering the \emph{classical} trajectory of the particle on the fixed background. This approximation is here valid if the particle can be considered point-like with respect to the Schwarzschild radius.
To calculate $\Gamma$ one has to do the following:
\begin{enumerate}
	\item  to solve explicitly the equation of motion $(t(\sigma),r(\sigma))$ in terms of a parameter (without loss of generality we can take $\sigma=r$); then to write the Lagrangian and thus the action \emph{through this parameter};
 \item to calculate the imaginary part of $\Delta S\equiv S_{AB}$ acquired in the extended complex plane $r$ in the path which goes from $A$ to $B$, the points of creation of the particle and of the antiparticle.
\end{enumerate}
For the Hawking radiation $A=r_H-0$ and $B=r_H+0$, then one has to calculate the half\footnote{Because we need just the path from $A$ to $B$.} of the Cauchy formula centered on the Schwarzschild radius $r_H$,
\[\Delta S=\frac{1}{2}\oint_{r_H} \mathcal L(r) dr \ .
\]
 This integral can give a non-null result only if the Lagrangian has a pole on $r_H$. We have seen that such pole on the horizon does not exist.
\section*{Appendix 2: Photons in Painlev$\acute{e}$ coordinates}
In the formulas (\ref{P.4}) and (\ref{P.5}) it is possible to take also the massless limit, considering which, for the out-going waves\footnote{The in-going ones have no poles at all.}, one finds:
\begin{eqnarray}
\texttt{Im}[\Delta W_P(r)]=\frac{1}{2}\texttt{Im}\oint \frac{-E}{\la- 1}\left(\frac{-r_H}{\la^3}\right)d\la=4\pi ME  \qquad \textit{(Parikh-Wilczek)} \label{A2.3} \\
\texttt{Im}[-E\Delta t_P(r)]=\frac{1}{2}\texttt{Im}\oint \frac{E}{\la- 1}\left(\frac{-r_H}{\la^3}\right)d\la=-4\pi ME \qquad \textit{(overlooked)} \label{A2.4} \ ,
\end{eqnarray}
and thus $\texttt{Im}(\Delta S)=0$. In Ref.\cite{PW00} the authors take into account only the contribution of (\ref{A2.3}); even if they considered also the back-reaction, while here we are ignoring this problem (i.e. we are considering the limit in which particle energy is much less than the Schwarzschild mass), the back-reaction considerations should not justify the omission of the temporal contribution (\ref{A2.4}).

\end{document}